# Impact on the Usage of Wireless Sensor Networks in Healthcare Sector

**Ahsan Humayun**
Lecturer
Department of
Computer Science
University of Central Punjab,
Faisalabad Campus, Pakistan

**Muneeb Niaz**
Lecturer
Department of
Computer Science
University of Central Punjab,
Faisalabad Campus, Pakistan

**Muhammad Umar**
Lecturer
Department of
Computer Science
University of Central Punjab,
Faisalabad Campus, Pakistan

**Muhammad Mujahid**
Lecturer
Department of
Computer Science
University of Central Punjab,
Faisalabad Campus, Pakistan

**Abstract**
Recent advancement in the wireless sensor networks has provided a platform to numerous applications in healthcare sector. It has become an active research area due to its large scale potential. This research focuses on the application areas of wireless sensor networks specifically in the healthcare sector. In this work, we have tried to explain the different challenges faced by the WSN's in order to implement them. The different pros and cons of the WSN's in healthcare sector are also discussed. Some important parameters which can be used to evaluate the performance of the wireless sensor networks are also presented in this work. Wireless sensor networks have a tremendous future and it should be taken at its earliest because of the significant importance of the healthcare issues.

## 1. Introduction

Wireless sensor networks consist of different sensor nodes that work collectively for a specific task. These nodes work within different type of environment with different conditions and circumstances. Basically there are two types of Wireless sensor networks: Structured and Unstructured. Initially Wireless sensor networks were only used for the military purposes for different purposes like monitoring and defense purpose. Now a day's Wireless sensor networks have a wide range of applications in our daily life routine. Now they have an emerging trend in our livings. They are used in many areas like healthcare, entertainment, industry, businesses and many more. In the recent years, Wireless sensor networks have progress a lot in the field of healthcare. They have wide application area in different medical fields' i.e. Patient monitoring automated medicine alert and diagnose of different disease and their procurement and many more [1].
Wireless sensor networks are quite different from traditional networks like WLAN and ad hoc networks. These types of networks just work to provide you the Quality of Service (QoS), while on the other hand the wireless sensor networks works in critical situations against some sort of unknown operations. For example, patients in which the wireless sensors are embedded, they don't need any kind of physician. In healthcare sector, now it's quite difficult to handle the patient status manually, now the WSN's have the potential to replace the manual work of monitoring and hence we can save our resources and time as well. Wireless sensor networks in healthcare have now become a hot research area, different applications with different scenarios have being researched like Heartbeat monitoring, blood pressure monitoring, body locations and positions etc. Due to this now the patient has ease to do his work freely and live like a normal person. He can easily perform his daily activities and hence at the same time he is being monitored for any issue regarding his health [2].

Wireless sensor networks are of different types and they are used on the basis of physical conditions, environment and conditions. Terrestrial wireless sensor networks are consisting of numerous sensor nodes which are deployed in a specific area for a specific purpose. Underground wireless senor networks are used to monitor the underground conditions and they are used for special purpose like searching oil resources and coal resources. They are deployed underground. Underwater wireless sensor networks are used to monitor the underwater activities. They are mainly used for the underwater surveillance for safety purposes. Multimedia wireless sensor networks are used to monitor the audio video or images. They are used in numerous applications. They are deployed with microphone and cameras. Mobile wireless sensor networks are used for monitoring the movements or locations of body in a physical environment. Mobile WSN's have features to sense different types of movements of objects [3].

## 2. Application Area of WSN in Healthcare Sector

Healthcare is always an important and critical concern as it is a matter of human livings. It has always a great importance in each aspect of field. There is a huge application area of WSN in healthcare sector. In this work we have tried to summarize the all applications area of WSN. We have categorized the applications of WSN in healthcare sector in different categories.

**Daily living Monitoring:** In these types of applications, different activities of patients like eating, sleeping, communicating with others is monitored. Basically RFID tags are used to accomplish these kinds of activities. These tags are embedded with the different objects of the user (keys, toothbrush, and glass) so that their activities can be monitored. It retrieves all the information automatically and present in graphical form.

**Movement Detection:** These types of applications are specifically used for aged people to monitor their movements.

**Location Tracking:** These types of applications are used to track down the physical locations of a person. They can be used both inside and outside, depending upon the conditions and the requirements. They are used to increase the context awareness [4].

**Intake of medication Monitoring:** Medication monitoring applications are very critical, they determine the time of medications for a patient and reminds him to take his medication, or if it embedded then it automatically injected the medication into the patient body.

**Patient status Monitoring:** These types of applications are widely used to monitor the status of patient either in hospital or home. They intimate the doctors if there is a critical situation with a patient. They monitor the heart rate, oxygen rate, blood pressure, body temperature and other information about a patient [5].

**Cancer Detection:** Cancer is now a day's considered as the most increasing disease and it has a major concern in health care issues. There are lots of methods to cure cancer, but it is quite difficult to monitor the continuous condition or status of a patient suffering from cancer. Results of different studies depicts that the cells affected by the cancer produce nitric oxide which cause harm and make a tumor. In order to monitor those cells, a special sensor which can specifically sense those cells are used and they are embedded inside the body of the patient. They monitor the locations of those suspected cells in the body and sends information in the form of report [6].

**Diabetes Detection:** Diabetes is another one of the rapidly increasing disease in the world. A lot of work is done on controlling the diabetes of the patients. Different systems on massive scale have been developed to control the diabetic conditions of the patients. With the help of wireless sensor networks we can control the diabetes of the patients. Special kind of sensor nodes are used which can regulate the insulin doze and it can calculate automatically the glucose rate inside the blood. For this purpose a special sensor node which is also known as biosensor is used and planted inside the human body [7].

**Asthma Detection**: Millions of people are now a day's suffering from asthma. It is also a fast growing disease due to increasing level of pollution in the environment. To overcome the asthma problem wireless sensor networks have a great impact, a special sensor nodes are used for this purpose which can detect the allergic agents found in the environment and hence they automatically and periodically update the doctor about the health status of the patient suffering from asthma. It detects and predicts the air and its condition and also alarm the patient if it forecast the air is polluted [8].

**Medical Accidents Prevention:** The most important issue in the healthcare sector is the human errors. Many of the accidents are caused due to the human faults. Sometimes the patient is not precisely checked by the doctor and hence the prescription of the doctor goes wrong which can lead death of the patient. In order to prevent such kind of accidents wireless sensor networks are used. According to a survey almost 98000 patients die every year due to such kind of errors**.**

**Strokes Prevention** Strokes are very serious and recovery from heart strokes is very difficult and crucial. It needs proper attention and consideration of doctor about the patient. In this regard, wireless sensor networks are used on a large scale to avoid such kind of activities. Now a day's wearable sensors are available which monitors the different parameters of the body and automatically update the data after a careful assessment and sends it to physician [9].

**Heart Rate Monitoring:** Many of the people die due to the heart attacks and heart failures, according to a survey almost 40 to 45 % people died due to heart attacks every year. Wireless sensor networks have progressed much to avoid the heart attacks. Different wrist watches have been developed by using the smart sensors with electrodes that take continuous care of the patient and regulate its body parameters and rapidly initiate the information in case of emergency [10].

## 3. Applications of WSN in Healthcare Sector

As healthcare is one of the most researched area almost in every field, so regarding the wireless sensor networks a number of real time applications are also developed, to address the different kind of healthcare issues. In this research, we have focus to provide a brief description of real time applications of wireless sensor networks [4].

**MobiHealth** is a mobile healthcare application which allows the patient to be mobile every time and they are being monitored continuously by using GRPS network.

**Vital Jacket** is also a smart mobile device, which is used as a wearable jacket used to monitor the heartbeat of the patient. It continuously measures the heart rate after a regular interval and updates the report of the patient to its consultant. It has alarming sensors which in case of emergency rapidly informs [11].

**AlarmNet** is also a mobile device, consisting of different wireless sensor networks. They sense the heart rate and oxygen rate, and alarm in case of any emergency.

**LifeGuard** was first developed for the astronauts. The basic purpose of this application is to monitor the blood pressure and pulse beat.

**CodeBlue** was a large scale project, and it provides a framework in providing rapid response against the disaster situations. It allows the quick remote monitoring and tracing the patient's situation.

**LifeShirt** is also a wearable shirt device, and is available commercially. It consists of different sensors and data recorders. It is a complete package including recording software which update regularly the patient status like its blood pressure, heart beat rate and pulse rate etc [12].

**SleepApnea** is very useful mobile device which is used to monitor the different parameters of the patient in night while he is sleeping, it monitors the blood circulation, the oxygen intake, breathing and the heart rate etc. it provides a rapid response in case of any casualty.

## 4. WSN Challenges In Healthcare Sector

**Security:** Wireless sensor networks have always been facing a very challenging issue of security due to its wireless nature. There are serious threats which can lead to severe security issues. For example in some of the case tracking the location by using the wireless sensor network can lead to serious consequences. It can lead to the negative usage of the WSN. A WSN should be smart enough that it can keep the data of the patient secure. It ensures the patient that the system is trustworthy and reliable and it will keep its data secure [13].

**Privacy** has always been a top most priority concern in wireless sensor networks with respect to healthcare issues. Patients are very much conscious about their information and their personal reports. It can be a very crucial threat towards the privacy issue of a patient. Individuals are very much careful about their data like where their data is stored, where and how it is used and to whom it is available. Another big privacy concern is regarding the sensor nodes which are planted inside the human body. Patients have psychological pressure and sometimes they feel that their privacy is no more with them [14].

**Power** is now a day's considered as a top most priority challenge, as WSN although need less power but they need continuous power in order to give good performance

**Continuous Performance** is one of the major challenges faced by WSN applications, in case of not responding regularly, it could be very dangerous and due to its sensitivity in healthcare issues it can lead to critical consequences [15].

**Robustness** is also a major concern of wireless sensor networks. Sensor nodes should ensure their long running time and performance regardless of the environment and the conditions they are facing.

**Data Synchronization** is another crucial challenge faced by the WSN applications. WSN must ensure that the correct data at the correct time is being sent and hence it must ensure the integrity of the patient's data [15].

## 5. Performance Parameters of WSN

There are different parameters on the basis of which we can evaluate the performance of wireless sensor networks. Some of the parameters have presented in this work which are of high importance [16].

**Network Lifetime** is considered as one of the important parameter, as the wireless sensor networks are battery operated, so it is necessary that they provide a longer lifetime.

**Coverage** is another critical parameter to evaluate the performance of wireless sensor networks, while using in some outdoor location, coverage of WSN should be reliable [17].

**Response Time** is another top priority parameter in order to measure the performance of the wireless sensor networks; it needs proper attention as it could be very critical. The response time of the WSN should be rapid and quick. And it should also ensure the accuracy of the data at the correct time.

**Deployment Ease** is also an essential parameter considered for the wireless sensor networks, the medical staff should be aware of the deployed locations of the sensor nodes, so that in case of any emergency they can be replace it easily or manage it easily [18].

## 6. Advantages & Disadvantages

The major advantage of wireless sensor network is that it overcomes the drawbacks of wired sensor networks. In wired sensor networks a hectic situation is created due to wires and it creates complications, and hence patient is restricted. WSN ensure the mobility of the patient as well as it provides ease for the hospital staff to manage [19].

The major disadvantage of wireless sensor networks is that when it is planted inside the human body, patients feels insecure. They feel that their privacy is revealed, and they have psychological pressure on their minds. Another disadvantage of using wireless sensor networks is that the cost maintaining these sensor nodes is a big expenditure for the patient because sensor nodes are expensive [20].

## 7. Conclusion & Future Directions

In this research, we have provided an impact on usage of wireless senor networks specifically in healthcare sector. Along this we have provided the different challenges faced by WSN, its advantages and disadvantages. We believe that the wireless sensor networks will have a large impact in future in the healthcare sector. The rapid increase in usages of wireless sensor networks has predicted that these smart sensors will be an integrated part of our daily life, as it already has a tremendous effect on our human life. There is credible future of self organizing WSN in healthcare sector. Wireless sensor networks can improve the systems like in-home assistance and smart nursing. Patient can ensure their privacy while remaining at home and healthcare services will be provided to them at their door step.